\documentstyle[12pt]{article}
\begin{document}

\begin{flushright}
UT-Komaba 96-8
\end{flushright}

%%%%%%%%%%%%%%%%%%%%%%%%%%%%%%%%%%%%%%%%%%%%%%%%%%%%%%%%%%%%%%%%%%%%%%%%

\begin{center} 
{\Large{\bf    
 Particle-Flux Separation of Electrons 
\vskip -0.1in
in the Half-Filled Landau Level: 
\vskip  0.1in
- Chargeon-Fluxon Approach -  
}}  
\vskip 0.5in
  {\Large  Ikuo Ichinose$^{\star}$\footnote{e-mail address:
 ikuo@hep1.c.u-tokyo.ac.jp} and 
Tetsuo Matsui$^{\dagger}$\footnote{e-mail address:
 matsui@phys.kindai.ac.jp} }
\vskip 0.2in
 $^{\star}$Institute of Physics, University of Tokyo, Komaba, Tokyo, 153 Japan\\
 $^{\dagger}$Department of Physics, Kinki University, Higashi-Osaka,  577 Japan
\end{center}
\vskip 0.2in
\begin{center} 
\begin{bf}
Abstract
\end{bf}
\end{center}
We have previously studied the  
phase structure at finite temperatures of the Chern-Simons (CS) gauge theory coupled with fermions by using lattice  gauge theory.
In this paper, we formulate the ``chargeon-fluxon" representation of electrons and use it to reinvestigate the phenomenon of
particle-flux separation (PFS) of electrons in the half-filled Landau level. 
We  start with a lattice system of fermions interacting with a CS
gauge field, and  
introduce two slave operators named chargeon and fluxon that carry the CS
charge and flux, respectively.
The original fermion, the composite fermion of Jain,
 is a composite of 
a chargeon and a fluxon. 
 We further rewrite the model by introducing an auxiliary link field,
the phase of which  behaves as a gauge field gluing chargeons
and fluxons. Then we study a confinement-deconfinement transition of that
gauge field by using  the theory of separation phenomena as in the
previous paper.
The residual four-fermi interactions   play an important role to determine
the critical temperature $T_{\rm PFS}$, below which the PFS takes place.
The new representation has some advantages; (1) It allows a  
field-theoretical description also for the flux degrees of freedom.
(2) It
has a close resemblance to the  slave-boson or slave-fermion
representations of the t-J model of high-Tc superconductors 
in which an electron is a composite
of a holon and a spinon. This   point opens a way to understand
the two typical separation phenomena in strongly-correlated electron systems
in a general and common setting.

\newpage
%%%%%%%%%%%%%%%%%%%%%%%%%%%%%%%%%%%%%%%%%%%%
\setcounter{footnote}{0}
\section{Introduction}
In the last several years, it has been recognized that the gauge theory
plays an essentially important role in some topics of condensed
matter physics.
Especially, for the fractional quantum Hall effect (FQHE) at
the    filling  factor 
$\nu=\frac{1}{2n+1} \; (n= 1,2,..)$, a Ginzburg-Landau
(GL) theory was proposed \cite{Girvin}, which is a gauge theory of a 
Chern-Simons (CS) gauge field  interacting with so-called bosonized electrons.
A FQH state is characterized as a Bose-condensation   of these bosonized
electrons.

Another important idea for the FQHE was proposed by Jain \cite{Jain}, that is, the
composite-fermion (CF) approach.
Jain asserted that the quasi-excitations in the half-filled Landau level
are fermions which he called  CF's; A CF  is nothing but a composite of
an electron and  two solenoidal flux quanta. 
The FQH states observed at a sequence of $\nu ={p\over 2p\pm1}$   $(p =1,2,..)$, are understood
as a result of the Landau-level quantization of {\em composite fermions},
i.e., integer QHE of CF's. 
The essential assumption of the CF approach at $\nu = 1/2$ is that two units 
of fluxes attaching to each CF  to form electrons 
cancel the  external magnetic field on the average and 
fluctuations of fluxes around the mean value behave almost independently
of the  locations (densities) of  CF's. 
Various experiments \cite{exper} and also numerical calculations \cite{RR} support, or at least
are consistent with, the idea of  CF.

The CS gauge theory is suitable for describing   CF's.
In the CS description, the above assumption means that the CS local
constraint, which connects  fluctuations of fluxes with the density of CF's
at each spatial point,
becomes irrelevent at low energies.
We called this phenomenon ``particle-flux separation 
(PFS)",\cite{IMO} because this
bears close resemblance to the charge-spin separation (CSS) in the strongly-correlated
electron systems for   high-Tc superconductivity \cite{Anderson}.
It is naturally expected that the repulsive interactions among
 electrons play  an essential role for the stability of CF's.

In the previous paper \cite{IMPFS}, we studied the PFS in the framework of   lattice
formulation of  CS gauge theory. 
We argued  that the PFS takes place {\it below}
 some critical temperature $T_{\rm PFS}$. 
(In Ref.\cite{IMPFS} we wrote it $T_{\rm CD}$.)
We applied the method also  to the quantum spin models and got the 
gauge-theoretical interpretation of each possible phases.

In the present paper, we shall address the same problem but with a different formalism.
The present method is closely related in its spirit with the slave-boson or slave-fermion formalism of the t-J model.
There the original electron operator is expressed in terms of a bilinear form
of spinon and holon operators which satisfy a local constraint.
If the CSS takes place, spinons and holons move almost freely as quasi-excitations.\footnote{Before our study of PFS \cite{IMPFS}, 
we  studied the CSS  \cite{IMCSS} in the framework of   gauge 
theory, where the gauge field glues  spinons
and   holons. Among other things, we showed that the CSS can be 
understood as the  deconfinement phenomenon of this gauge dynamics,
and calculated the critical temperature $T_{\rm CSS}$ below which 
the CSS takes place.} 
In the present formulation of the CS gauge theory of fermions, 
we   introduce yet another set of two slave
operators to express the fermion operator, the original CF operator, as a  bilinear form of them.
We call them  chargeon and fluxon operators since they carry charges and 
fluxes, respectively.\footnote{These charges and fluxes are of 
the CS gauge field and {\it not} of the usual electromagnetism.}
In the CS gauge theory of fermions, fermions move in a statistical 
magnetic field,
and that statistical magnetic field is made of certain amount of  flux quanta attaching   to each fermion due to the CS constraint.
Therefore, a fermion carries  a magnetic flux as well as a charge for the CS vector potential.
In order to discuss the PFS, it seems natural to introduce corresponding 
operator for each property, the CS  charges
and the CS magnetic fluxes, i.e., the chargeon and the fluxon operators.
The PFS is understood as a deconfinement phenomenon of  chargeons and
fluxons. Furthermore, introduction of fluxon operators makes it possible to describe possible excitations of magnetic fluxes as independent  quasi-excitations. 

This paper is organized as follows.
In Sect.2, we introduce a model of the CS gauge theory 
interacting with fermions,
and explain its relationship with electrons in the half-filled 
Landau level.
The chargeons and fluxons are introduced similarly as the spinons and  holons in the t-J model.
We further rewrite the model by a Hubbard-Stratonovich transformation.
The auxiliary link field introduced there is regarded as a 
gauge field which glues  a chargeon  and a fluxon.
In Sect.3, we derive an effective action for that gauge field.
It is explained that the deconfinement phase of that gauge field is the phase of  PFS. The effective gauge-coupling constant is explicitly calculated
 as a function of the temperature $T$
and the concentration of fermions.
In Sect.4, we study the phase structure of the effective gauge theory,
showing that the PFS takes place {\it below} certain critical temperature
$T_{\rm PFS}$. The value of $T_{\rm PFS}$ is  slightly different from that
given in Ref.\cite{IMPFS} reflecting the different representations.
We see that the Coulombic repulsion between electromagnetic (EM) charges 
is very important for PFS, which supports the intuitive physical 
expectation for the stability problem of CF's.
Section 5 is devoted for conclusion.

%%%%%%%%%%%%%%%%%%%%%%%%%%%%%%%%%%%%%%%%%%%%%%%%%
\setcounter{equation}{0}
\section{Electrons in the half-filled Landau level and the chargeon
and fluxon operators}

\subsection{Model}

Let us start with a model of CS gauge theory coupled with spinless nonrelativistic 
fermions $\psi_x$ on a 2-dimensional  spatial lattice. These fermions
are identified with the CF's at the half filling. We employ the imaginary-time
formalism at finite $T$. Its Lagrangian  is given by
\begin{eqnarray}
L&=&-\sum_x \psi^{\dagger}_x(\partial_{\tau}-iA^{\rm CS}_0-\mu_c)\psi_x
+{1\over 2m}\sum_{x,j}(\psi^{\dagger}_{x+j}
e^{i(A^{\rm CS}_{xj}-e A^{\rm ex}_{xj})}\psi_x+\mbox{H.c.})   \nonumber  \\
&&-{i\over 2\pi q}\sum_{x,\mu\nu\lambda} 
\epsilon_{\mu\nu\lambda}A^{\rm CS}_{x\mu}F^{CS}_{x\nu\lambda} 
+ L_{\rm int}(\psi^{\dagger}_x\psi_x),
\label{model1}
\end{eqnarray}
where the Grassmann number $\psi_x (\tau)$ and
the CS gauge potential $A^{\rm CS}_{x\mu}(\tau)$ are functions
 of the imaginary time  $\tau$, where $0 < \tau < \beta \equiv 1/T$.  
$x$ denotes the lattice sites, $q$ is a parameter of the model,
Greek indices $\mu, \; \nu,...$  
take $ 0,1,2$ and denote 
the directions ($0$; imaginary-time $\tau$, $1,2$; spatial), and 
$j  = 1,2$ (and $i,k$ appearing later) denotes the spatial directions.
The field strength $F_{\mu\nu}$ and the CS magnetic field $B^{\rm CS}_x$ 
are given by
\begin{eqnarray}
F_{xij}&=&\nabla_i A^{\rm CS}_{xj}-\nabla_j A^{\rm CS}_{xi},  \nonumber  \\
B^{\rm CS}_x & \equiv & F_{x12} \nonumber  \\
F_{x0i}&=&\partial_{\tau}A^{\rm CS}_{xi}-\nabla_iA^{\rm CS}_{x0},
\label{strength}
\end{eqnarray}
where $\nabla_i$ is the lattice difference operator.
We have also introduced the external electromagnetic 
field $A^{\rm ex}_{xi}$ in the model.
$L_{\rm int}(\psi^{\dagger}_x\psi_x)$ represents the residual interactions 
among fermions.   

Let us see that the electron system in a uniform magnetic field $B^{\rm ex}$
at Landau filling factor $\nu={1\over 2}$ (or more generally $\nu=1/(2n)$
with a positive integer $n$) is described by the above model.
The Hamiltonian of the electron system is given by 
\begin{equation}
H_e=-{1\over 2m}\sum\Big( C^{\dagger}_{ x+j}e^{-ieA^{\rm ex}_{xj}}C_{x}
+\mbox{H.c.}\Big) - L_{\rm int}(\{C^{\dagger}_{ x}C_{ x}\}),
\label{He}
\end{equation}
where $C_{x}$ is the polarized electron annihilation operator at site $x$,
and the vector potential $A^{\rm ex}_{xi}$ describes $B^{\rm ex}$,
$$
B^{\rm ex}=\epsilon_{ij}\nabla_iA^{\rm ex}_{xj}.
$$
$L_{\rm int}(\{C^{\dagger}_{ x}C_{ x}\})$ represents interactions 
among electrons.
The lattice spacing $a$, which is often set to unity, is identified with the magnetic length.  

To this end, we first differentiate $L$ of  (\ref{model1}) w.r.t.  $A^{\rm CS}_{x0}$  to obtain the equation of motion, 
\begin{eqnarray}
B^{\rm CS}_x &=&2\pi q \hat{\rho}_x,
\nonumber\\
\hat{\rho}_x &\equiv& \psi^{\dagger}_x \psi_x,
\label{BCS}
\end{eqnarray}
which is the well-known  CS constraint showing that the CS fluxes
of $q $ units are attached to each fermion $\psi^{\dagger}_x$.
Eq.(\ref{BCS}) can be solved  in the transverse gauge $\nabla_i A^{\rm CS}_{xi} =0$ as 
\begin{eqnarray}
A^{\rm CS}_{xi}&=&2\pi q\epsilon_{ij}\sum_y\nabla_jG(x,y)\hat{\rho}_y \nonumber\\
&=& \sum_y\nabla_i\theta(x-y)\hat{\rho}_y ,
\label{vecpCS}
\end{eqnarray}
where $G(x,y)$ is the 2-dimensional Green function, and  
$\theta(x)$ is the multi-valued angle function on a lattice with $\theta (0)=0$.
The corresponding Hamiltonian is  given by the standard procedure as
\begin{equation}
H_{\psi}=-{1\over 2m}\sum\Big( \psi^{\dagger}_{x+j}e^{i(A^{\rm CS}_{xj}-eA^{\rm ex}_{xj})}\psi_{x}
+\mbox{H.c.}\Big) - L_{\rm int}(\{\psi^{\dagger}_{x}\psi_{x}\}),
\label{HCF}
\end{equation}
where  $A^{\rm CS}_{xi}$ is given by (\ref{vecpCS}). 

Next, let us introduce the operator $C_x$ as
\begin{equation}
C_{ x} \equiv \exp \Big[ i q\sum_y \theta(x-y)\hat{\rho}_y\Big] \psi_x,
\label{CF}
\end{equation}
where 
\begin{equation}
\hat{\rho}_x=C^{\dagger}_{ x}C_{ x}=\psi^{\dagger}_x\psi_x.
\label{rho}
\end{equation}
For the half-filled Landau level, we put the parameter $q=2$ in (\ref{CF}).
Then one can check that $C_x$ satisfy the canonical anticommutation relations
(CACR) for fermions.
By substituting (\ref{CF}) into (\ref{HCF}), we reach the Hamiltonian 
of electron system (\ref{He}).

As the filling factor $\nu$ is given by 
$\nu=2\pi \rho/eB^{\rm ex}$, where $\rho$ is 
average density of electrons or $\psi_x$'s, 
$\rho=\langle \hat{\rho}_x\rangle$,
we have
\begin{equation}
\langle B^{\rm CS}\rangle=eB^{\rm ex},
\label{BB}
\end{equation}
for $\nu=1/q$ from (\ref{BB}).
Therefore, the external magnetic field is cancelled out 
on the average by the CS magnetic field.

%%%%%%%%%%%%%%%%%%%%%%%%%%%%%%%%%%%%%%%%%%%%%%%%%%%%%%%%%%%%%%
\subsection{Chargeon and fluxon}

From the Hamiltonian (\ref{HCF}) and the CS constraint (\ref{BCS}), 
it is obvious that 
a $\psi_x$ quantum carries $q$ magnetic flux quanta,
and, at the same time, interacts  minimally  with the CS magnetic field.
It plays  a dual role  of  matter field and source of the force field. 
As explained in the introduction, the essential assumption of the CF approach
to the FQHE is that the CS constraint becomes irrelevant at low energies and
correlations between charge  and flux degrees of freedom become weak, i.e., the PFS takes place for quasi-excitaions.

To describe this possibility, let us introduce the chargeon operator
$\eta_x$ and the fluxon operator $\phi_x$, and express the fermion 
operator $\psi_x$ as  
\begin{equation}
\psi_x=\phi_x\eta_x.
\label{bilinear}
\end{equation}
In order that the operator $\psi_x$  satisies the fermionic CACR, 
i.e., $\{ \psi_x, \psi^{\dagger}_y \}=\delta_{xy}$, 
we assign statistics for the chargeon and the fluxon 
as follows,\footnote{Of course 
we can assign the alternative statistics; $\eta_x$: hard-core boson, 
$\phi_x$: fermion.  
Final result for the PFS does not depends on which assignment is employed.
However, this assignment is more suitable as the chargeons behave as 
 CF's in the state of  PFS. See later discussion.}
\begin{eqnarray}
{\rm chargeon}\ \ \eta_x &:& \mbox{fermion},  \nonumber   \\
{\rm fluxon} \ \ \phi_x &:& \mbox{hard-core boson},
\label{statistics}
\end{eqnarray}
and impose that they have to satisfy the following local constraint;
\begin{equation}
\eta_x^{\dagger}\eta_x = \phi_x^{\dagger}\phi_x .
\label{cons}
\end{equation}
By the term hard-core bosons, we mean  that $\phi_x$  and $\phi_y$
 satisfy the usual canonical commutation relation (CCR) of 
bosons,  $[\phi_x , \phi_y] =  [\phi_x , \phi^{\dagger}_y]=0,$ 
 for  $x \neq y$, and satisfy
 CACR, $\{\phi_x , \phi_x\} =0, \{\phi_x , \phi^{\dagger}_x\}=1,$ 
 for each $x$.\footnote{ One may check easily that still another 
 assignment for $\phi_x$,
 the CCR (for all $x$ and $y$) {\it of canonical bosons  instead   
of hard-core bosons}, together with (\ref{cons}), works well
 ($\eta_x$ is kept fermion). Since
the   results obtained in later Sections are   unchanged,
we use (\ref{statistics}) for definiteness.}
We also assign that $ \eta_x$'s and $\phi_x$'s  commute each other.
The physical meaning of the above constraint (\ref{cons}) is understood
as follows.
For each site $x$, there are two physical states, no fermion state 
$|0 \rangle 
 (\psi_x |0 \rangle  =0)$
and one fermion state $|1\rangle   \equiv \psi^{\dagger}_x |0\rangle $.
They are desscribed   using  $\eta_x$  and $\phi_x$ and their vacuum state
$|V\rangle , \eta_x |V\rangle  = \phi_x |V\rangle  =0$, as 
\begin{eqnarray}
|0\rangle  &=& | V\rangle , \nonumber\\ 
|1\rangle  &=& \eta^{\dagger}_x \phi^{\dagger}_x|V\rangle .
\end{eqnarray}
Actually, the constraint (\ref{cons}) allows for only these states.
We should be careful for ``defining" operators like the number operator 
$\psi^{\dagger}_x\psi_x$.
Its consistent definition with respect to (\ref{cons}) is given as\footnote{Remark that  
single operators like $\eta_x$ or $\phi_x$ do  {\em not} commute with 
the constraint (\ref{cons}).} 
\begin{eqnarray}
\psi^{\dagger}_x\psi_x &=&\eta^{\dagger}_x\eta_x\phi^{\dagger}_x\phi_x \nonumber\\
&=& \eta^{\dagger}_x\eta_x \eta^{\dagger}_x\eta_x \nonumber\\
&=&\eta^{\dagger}_x\eta_x   =\phi^{\dagger}_x\phi_x.
\end{eqnarray}
To implement the meaning of flux annihilation operator to $\phi_x$,
we write the CS constraint in the form,
\begin{eqnarray}
B^{\rm CS}_x &=& 2\pi q    \phi^{\dagger}_x \phi_x,
\label{BCS2}
\end{eqnarray}
which manifestly shows that a fluxon carries $q$ units of CS flux quanta. We respect this equation faithfully irrespective of whether the PFS occurs or not. 
The very condition for PFS is that the chargeon-fluxon constraint 
(\ref{cons}) becomes irrelevant to quasi-excitations, as we argue just below.  
Since our strategy is to prepare two operators, each by each for   fluxes and 
CS charges, we assign so that a chergeon carries one unit of CS charge. 
 These properties and other assignment are summerized in Table 1.

In terms of these operators, let us express the Hamiltonian.
By substituting (\ref{bilinear}) into (\ref{HCF}), we have
\begin{eqnarray}
H_{\eta\phi}&=&-{1\over 2m}\sum \Big(\eta^{\dagger}_{x+j}\phi^{\dagger}_{x+j}
W_{x+j}W^{\dagger}_x\phi_x\eta_x+ \mbox{H.c.}\Big) -      
 L_{\rm int}(\{\eta^{\dagger}_x\eta_x,
\phi^{\dagger}_x\phi_x\})  \nonumber   \\
&&\; -\sum (\mu_{\eta}\eta^{\dagger}_x\eta_x+\mu_{\phi}\phi^{\dagger}_x\phi_x)
- \sum \lambda_x(\eta^{\dagger}_x\eta_x-\phi^{\dagger}_x\phi_x),
\label{HC-F}
\end{eqnarray}
where
\begin{equation}
W_x=\exp \Big[iq\sum_y\theta(x-y)(\phi^{\dagger}_y\phi_y-\rho)\Big].
\label{W}
\end{equation}
Here we used the expression,
\begin{eqnarray}
e^{-ieA^{\rm ex}_{xj}} = \exp \Big[-iq\sum_y\theta(x-y)\rho \Big].
\end{eqnarray}
We have introduced Lagrange multiplier field $\lambda_x$ for the slave-particle
constraint (\ref{cons}), and  the chemical potentials $\mu_{\eta}$ and $\mu_{\phi}$ 
(chargeons and fluxons must have the same average density,
$\langle\eta^{\dagger}_x\eta_x\rangle=\langle\phi^{\dagger}_x\phi_x\rangle$).
This Hamiltonian (\ref{HC-F}) is invariant under the following local gauge
transformation;
\begin{equation}
(\eta_x,\phi_x) \rightarrow (e^{i\alpha_x}\eta_x, e^{-i\alpha_x}\phi_x),
\label{gaugetrf}
\end{equation}
whose origin is obvious from (\ref{bilinear}).

We shall investigate the possibility of   PFS.
In the state of PFS,    chargeons and fluxons are 
 {\it quasi-excitations} and move almost independently.
The constraint (\ref{cons}) is not faithfully respected
by them.\footnote{The slave-particle constraint (\ref{cons})
is surely satisfied by the original {\em bare} operators.
However, if the PFS takes place, the same constraint 
is not satisfied by the {\em asymptotic fields} 
for quasi-particles at low energies.
See Subsect.3.1 for more concrete discussion.}
We shall start with (\ref{BCS2}) instead of (\ref{BCS}), and 
investigate nature of the quasi-excitations.
This assertion is motivated by the success of the CF idea.
Similar (but more loose) approximation is recently used in
 the renormaization-group (RG)
study of CS gauge theories of nonrelativistic fermions \cite{RG}.
There, Coulombic-type  interaction term  between fermions  
is converted to a kinetic term
of the ``CS" gauge field through the CS constraint (\ref{BCS}) or
(\ref{BCS2}), 
and then the CS constraint is neglected in the RG transformation.
This approximation gives interesting and physically acceptable results,
though there appeared no solid justifications yet.
Strictly speaking, there is a problem that there seem to be 
 no {\it a priori} principles to fix the most suitable representation 
 of Hamiltonian to study quasi-excitations in the PFS state.
There are lots of representations that are all equivalent  under the CS
 constraint, but they become different right after relaxing the constraint.
Our chargeon-fluxon representation gives a partial answer to this problem;
prepare an  operator for each possible degree of freedom (charge, flux,..),
and then relax the constraint for these operators. In this sense, the present
approach is {\it different} from our previous analysis \cite{IMPFS} in which
we did not introduce fluxon operators explicitly. (We shall explain more on it
below.)

Before going into details of the analysis of PFS,
it may be instructive to compare the above expressions with those of the t-J model in the slave-boson or fermion formalism.
In the slave-boson formalism, the electron operator 
$C_{x \sigma}$ ($\sigma=\uparrow,
\downarrow$) is expressed as 
\begin{equation}   
C_{x \sigma}=b^{\dagger}_xf_{x \sigma},
\label{slaveboson}
\end{equation}
where $b_x$ is bosonic holon operator that carries charge $+e$, 
and $f_{x \sigma}$
is fermionic spinon operator that is charge-neutral and has a spin.
The local constraint is given by 
\begin{eqnarray}
b^{\dagger}_xb_x+\sum_{\sigma}f^{\dagger}_{x \sigma}f_{x \sigma}=1.
\end{eqnarray}
In the state of   CSS,  holons and  spinons move almost independently
with each other and they interact perturbatively with dynamical gauge fields
which are phase degrees of freedom of link mean fields \cite{IMCSS}.
Then the holons and  spinons appear as quasi-particles and they
do not satisfy the above local constraint.
The analogy becomes more striking if one makes a particle-anti-particle transformation,  $\phi_x \rightarrow 
\phi^{\dagger}_x$ in (\ref{bilinear}) and (\ref{cons}). Then they become
\begin{eqnarray}
&&\psi_x  = \phi^{\dagger}_x \eta_x, \nonumber\\
&& \phi^{\dagger}_x\phi_x +\eta^{\dagger}_x
\eta_x = 1.
\label{const2}
\end{eqnarray}
This implies  the  correspondences,  
\begin{eqnarray}
\psi_x \leftrightarrow C_{x\sigma}, \; \; \eta_x \leftrightarrow f_{x \sigma},
\; \; \phi_x \leftrightarrow b_x.
\label{coresp}
\end{eqnarray}

%%%%%%%%%%%%%%%%%%%%%%%%%%%%%%%%%%%%%%%%%%%%%%%%%%%%%

\subsection{Auxiliary gauge field, PFS, and CF's}

The partition function $Z \equiv {\rm Tr}\exp(-\beta H_{\eta\phi})$ for (\ref{HC-F}) is expressed
in path-integral formalism as 
\begin{eqnarray}
Z &=& \int[d\eta][d\phi]\exp(\int_{0}^{\beta} d\tau L_{\eta\phi}).
\end{eqnarray}
The Lagrangian is                
\begin{equation}  
L_{\eta\phi}=-\sum\eta^{\dagger}_x\partial_{\tau}\eta_x-
\sum\phi^{\dagger}_x\partial_{\tau}\phi_x-H_{\eta\phi}.
\label{Lag}
\end{equation} 
Here $\eta_x(\tau)$ is  a Grassmann number. The integration over the hard-core bosons $\phi_x$ needs a special treatment. We treat $\phi_x(\tau)$ as
a complex number. Its hard-core nature can be incorporated faithfully 
in the hopping expansion carried out  below. 	
Now let
us perform a Hubbard-Stratonovich transformation to introduce an auxiliary complex field $V_{xj}$ on  the  link $(x,x+j)$. Then we have
\begin{eqnarray}
Z &=& \int[d\eta][d\phi][dV]\exp( \int_{0}^{\beta} d\tau L_{\eta\phi V}),
\end{eqnarray}
and then the Lagrangian $L_{\eta\phi V}$ is given by 
\begin{eqnarray}
L_{\eta\phi V} &=&-\sum\eta^{\dagger}_x(\partial_{\tau}+i\lambda_x-\mu_{\eta})\eta_x
-\sum\phi^{\dagger}_x(\partial_{\tau}-i\lambda_x-\mu_{\phi})\phi_x  \nonumber \\
&&+\sum {1\over 2m}\Big[V_{xj}(\phi_{x+j}\phi^{\dagger}_x+
\eta^{\dagger}_{x+j}W_{x+j}W^{\dagger}_x\eta_x)+\mbox{H.c.}\Big] \nonumber\\
&&-\sum{1\over 2m}\Big(\phi^{\dagger}_{x+j}\phi_x\phi^{\dagger}_x
\phi_{x+j}\Big)
-\sum{1\over 2m}\Big(\eta^{\dagger}_{x+j}\eta_x\eta^{\dagger}_x\eta_{x+j}\Big)
\nonumber\\
&&-\sum {1\over 2m}|V_{xj}|^2
+ L_{\rm int}(\{\eta^{\dagger}_x\eta_x,\phi^{\dagger}_x\phi_x\}).
\label{LCF}
\end{eqnarray} 
This  Lagrangian (\ref{LCF}) clearly shows that a chargeon $\eta_x$
moves in the statistical magnetic field that is  generated by  fluxons 
(\ref{W}), as  we  expected. This is welcome since we are now free from
facing a complicated problem that a single field hops through an effective
 field generated via itself nonlocally.  
Furthermore, both chargeons and fluxons couple minimally to 
 the link field $V_{xj}$,  which gives rise to  attractive interactions 
between a chargeon $\eta_x$ and a fluxon $\phi_x$. It induces  
attractions also in the $\eta^{\dagger}_{x+j}-\eta_x$  
and $\phi^{\dagger}_{x+j}-\phi_x$ channels.
The four-Fermi $\eta^4$ term and the $\phi^4$ term in (\ref{LCF}) appear to cancel these residual ``gauge" interactions. See Table 1 for the charges that couple to the gauge field $V_{xj}$, which we denoted ``V charge".
In Fig.1 we illustrate the key concepts and objects appeared 
in each step to reach $L_{\eta\phi V}$ of (\ref{LCF}), such as  $C^{\dagger}_x, \psi^{\dagger}_x, \tilde{W}_x, \eta^{\dagger}_x, \phi^{\dagger}_x, W_x$.

As a model of the half-filled Landau level,
the interaction term of  chargeons and   fluxons, 
$L_{\rm int}(\{\eta^{\dagger}_x\eta_x,\phi^{\dagger}_x\phi_x\})$ in (\ref{LCF}),
is obtained from
the interactions between electrons in the following way.
For example, let us assume the following nearest-neighbor interactions;
\begin{eqnarray}
L_{\rm int}(\{C^{\dagger}_{ x}C_{ x}\})&=&
L_{\rm int}(\{\psi^{\dagger}_{x}\psi_{x}\}) \nonumber \\
&=&g\sum \psi^{\dagger}_{x+j}\psi_{x+j}\psi^{\dagger}_x\psi_x,
\label{Lint1}
\end{eqnarray}
where $g$ is the coupling constant.
Then from the chargeon-fluxon constraint (\ref{cons}) and  the Fermi statistics of $\eta_x$, it can be written as 
\begin{eqnarray}
\psi^{\dagger}_{x+j}\psi_{x+j}\psi^{\dagger}_x\psi_x&=&(\phi^{\dagger}_{x+j}
\phi_{x+j}\eta^{\dagger}_{x+j}\eta_{x+j})(\phi^{\dagger}_x
\phi_x\eta^{\dagger}_x\eta_x)  \nonumber \\
&=&\eta^{\dagger}_{x+j}\eta_{x+j}\eta^{\dagger}_x\eta_x \nonumber\\
&=&\phi^{\dagger}_{x+j}\phi_{x+j}\phi^{\dagger}_x\phi_x.
\label{Linter}
\end{eqnarray}
This leads us to generalize  the model by  introducing two coupling constants  
 $g_1$ and $g_2$ to write
\begin{equation}
L_{\rm int}(\{\eta^{\dagger}_x\eta_x,\phi^{\dagger}_x\phi_x\}) =g_1 \sum 
\eta^{\dagger}_{x+j}\eta_{x+j}\eta^{\dagger}_x\eta_x+g_2\sum
\phi^{\dagger}_{x+j}\phi_{x+j}\phi^{\dagger}_x\phi_x.
\label{inter}
\end{equation}
For example,  if the interaction (\ref{Lint1}) represents  Coulombic 
(but short-range) repulsion between
{\em the } EM {\em charges}, the parameters are 
$g_1 < 0$ and $g_2 = 0$ since we asign that
$\eta_x$ and $\phi_x$ carry the EM charge $-e$ and $0$,
 respectively (See Table 1).  
 
It is obvious that if the amplitude of link field $V_{xj}$ is novanishing, its phase degrees of freedom exist and behave as a gauge field, because under the gauge transformation (\ref{gaugetrf}), $V_{xj}$ transforms as
\begin{equation}
V_{xj}\rightarrow e^{i\alpha_{x+j}}V_{xj}e^{-i\alpha_x}.
\label{gaugetrfV}
\end{equation}
If the gauge dynamics of $V_{xj}$ is in the confinement phase, the gauge
field fluctuates largely, giving rise to the vanishing expectation value of 
$V_{xj}$. The  only charge-neutral
compounds with respect to (\ref{gaugetrf}) and (\ref{gaugetrfV})
appear as physical excitations, such as $\eta^{\dagger}_x\phi^{\dagger}_x$, 
which are nothing but the original fermions $\psi^{\dagger}_x$.
Due to the CS constraint (\ref{BCS}),  each of  these fermions necessarily accompany $q$ units of fluxes, forming the electrons for $q=2$. 
Therefore we conclude that the quasi-excitations at half filling  are the  original electrons when the gauge dynamics is realized in the confinement phase.
On the other hand, if it is in the deconfinement phase, gauge fluctuations are small and $V_{xj}$ develops a quasi-long-range order.
Therefore chargeons and  fluxons acquire their own  hopping amplitudes which are proportional to $\langle  V_{xj} \rangle \neq 0$, 
so behave as quasi-excitations. 
Especially, the chargeons $\eta_x$  describe 
 nothing but the weakly interacting CF's proposed by Jain.
This last point deserves more explanation.
In the literatures the word ``composite fermion" is used for almost free fermions which appear as a result of the cancellation between the external magnetic field and the average of the CS field with {\em neglecting} the CS constraint (\ref{BCS}). In this sense, the chargeons in our case can be regarded as CF's.
However, there is another possible definition of  CF's, that is, 
the field $\psi_x$ itself may describe  CF's. This interpretation has been  advocated in Ref.\cite{IMPFS}. There we worked directly with $\psi_x$ 
({\it without} further decomposition into chargeon and fluxon) and introduced the  same auxiliary link field $V_{xj}$. The hopping term in the action thus reads like
\begin{eqnarray}
\sum  \Big[V_{xj}(\psi^{\dagger}_{x+j}\psi_x+
 W^{\dagger}_{x+j}W_x)+\mbox{H.c.}\Big],
\end{eqnarray}
where the nonlocal operator $W_x$ is given by the same expression (\ref{W})
{\it but} with  $\psi^{\dagger}_x\psi_x$  for its source instead of $\phi^{\dagger}_x\phi_x$. We argued that the PFS takes place if the gauge dynamics is in the deconfinement phase, and the quasi-excitations are
$\psi_x$ particles, which, in turn, we interpreted as CF's.
The difference between these two candidates for CF's, $\eta_x$ and $\psi_x$, 
lies whether they have own CS fluxes or not. (See Table 1.)
However, this difference is subtle. Actually, at $T=0$, one may expect that
the boson field $\phi_x$ may Bose condense. Then, in the leading treatment, 
one can set $\phi_x$ as a constant $\phi_x = \sqrt{\rho}$, 
which washes out the difference of two operators since $\psi_x = \phi_x \eta_x
\propto \eta_x$.
 The difference may appear in the next order in the  small fluctuations 
$\phi_x = \sqrt{\rho} +\delta\phi_x$.
In Sect.4 we mention the perturbative analyses in the literature in connection with this point. However, we find 
{\it no} strong  reasons why one interpretation is better than the other.
They are physically equivalent in the leading (mean-field) treatment, and there appear no explicit quantitative comparisons of higher-order corrections. 
What we can say is that  the chargeon-fluxon approach manifests the separation of degrees of freedom, and opens a possibility to describe the dynamics of charges and fluxes on an equal footing.    We expect that it is certainly superior 
if the system supports nontrivial but quasi-local 
flux excitations  as quasi-excitations.

In the following section, we shall obtain an effective gauge theory of $V_{xj}$
from (\ref{LCF}) by intergrating out the fields $\eta_x$ and $\phi_x$.
To this end, we use the hopping expansion as  in the previous 
studies of the CSS and the PFS \cite{IMPFS,IMCSS}.

%%%%%%%%%%%%%%%%%%%%%%%%%%%%%%%%%%%%%%%%%%%%%%%%
\setcounter{equation}{0}
\section{Effective gauge theory}
In this section, we  derive an effective gauge theory of $V_{xj}$.
We first decompose $V_{xj}$ into its amplitude and phase variable,
\begin{equation}
V_{xj}=V_0\; U_{xj}, \; U_{xj}\in U(1).
\label{U1}
\end{equation}
As explained in the previous section, $U_{xj}$ behaves as a gauge field
and plays an important role to determine  the  spectrum of low-energy  
 excitations.    In Sect.3.1 we start with a   brief discussion  
 on the relationship between the local constraint and the gauge dynamics. 
Then we study the behavior of the amplitude $V_0$ as a function of $T$,
by using the hopping expansion w.r.t. $\eta_x$ and $\phi_x$ in Sect.3.2,
and by using the mean-field type calculation in Sect.3.3. In Sect.3.4,
we derive the effective action of $U_{xj}$ by the hopping expansion.
The usefulness of the hopping expansion in such a  situation has been
 explained in Ref.\cite{IMPFS,IMCSS}.

%%%%%%%%%%%%%%%%%%%%%%%%%%%%%%%%%%%%%%%%%%%%%%%%%
\subsection{Local constraint and gauge dynamics}
The effective action $A[V]$ is defined 
as\footnote{As stated before, this path-integral expression 
is rather formal for the hard-core boson $\phi_x$.
We shall use the knowledge of operator formalism for deriving  
propagators and calculating matrix elements.}
\begin{equation}
e^{A[V]}=\int [d\eta][d\phi] [d\lambda]
\exp \Big[\int^{\beta}_0d \tau L_{\eta\phi V}\Big].
\label{effA}
\end{equation}
In the hopping expansion to evaluate $A[V]$, one needs the  
propagators of $\eta_x$ and $\phi_x$, which are obtained from  ({\ref{LCF}) as
\begin{eqnarray}
&&\langle\eta_x(\tau_1)\eta^{\dagger}_y(\tau_2)\rangle=\delta_{xy}G_{\eta}
(\tau_1-\tau_2),     \nonumber   \\
&&G_{\eta}(\tau)={e^{\mu_{\eta}\tau} \over 1+e^{\beta\mu_{\eta}}}
[\theta(\tau)-e^{\beta\mu_{\eta}}\theta(-\tau)], \nonumber  \\
&&\rho={e^{\beta\mu_{\eta}} \over 1+e^{\beta\mu_{\eta}}}, \label{Geta}  \\
&&\langle\phi_x(\tau_1)\phi^{\dagger}_y(\tau_2)\rangle=\delta_{xy}G_{\phi}
(\tau_1-\tau_2),     \nonumber   \\
&&G_{\phi}(\tau)={e^{\mu_{\phi}\tau} \over 1+e^{\beta\mu_{\phi}}}
[\theta(\tau) - e^{\beta\mu_{\phi}}\theta(-\tau)], \nonumber  \\
&&\rho={e^{\beta\mu_{\phi}} \over 1+e^{\beta\mu_{\phi}}}.
\label{Gphi}  
\end{eqnarray}

In order to see how the chargeon-fluxon constraint (\ref{cons}) is affected
by the gauge dynamics, let us calculate the quadratic term of $\lambda_x$
in the leading order. 
From (\ref{LCF}), the contribution from   $\phi_x$ is estimated as follows;
\begin{eqnarray}
&&-\int d\tau_1d\tau_2d\tau_3d\tau_4\;\lambda_x(\tau_1)\lambda_{x+j}(\tau_3) 
\nonumber  \\
&& \; \; \times \;\langle \phi^{\dagger}_x(\tau_1)\phi_x(\tau_1)
\phi^{\dagger}_{x+j}(\tau_2)\phi_x(\tau_2)\phi^{\dagger}_{x+j}(\tau_3)\phi_{x+j}(\tau_3)
\phi^{\dagger}_x(\tau_4)\phi_{x+j}(\tau_4)\rangle  \nonumber  \\
&& \; \; \times \; \langle V^{\dagger}_{xj}(\tau_4)V_{xj}(\tau_2)\rangle  \nonumber  \\
&& \; \; =-\int d\tau_1d\tau_2d\tau_3d\tau_4\;\lambda_x(\tau_1)\lambda_{x+j}(\tau_3) 
\; \prod_i G_{\phi}(\tau_{i+1}-\tau_i)\cdot 
\langle V^{\dagger}_{xj}(\tau_4)V_{xj}(\tau_2)\rangle,
\label{lambda}
\end{eqnarray}
where $\tau_{i+4}=\tau_i$.
From (\ref{Gphi}) it is verified that the $\tau_i$-dependence of
$\prod_i G_{\phi}(\tau_{i+1}-\tau_i)$ (especially $\tau_2$ and $\tau_4$-dependence)
cancels with each other and only the $\theta$-functions remain.
Therefore, the relevant $\tau_i$-dependence of the integrand in 
(\ref{lambda}) may stem only from the correlation function of $V_{xj}$.

It is convenient to introduce Fourier decomposition of the gauge 
field $U_{xj}$,
\begin{eqnarray}
&& U_{xj}(\tau)=\sum_ne^{i\omega_n\tau}U_{xj,n},  \nonumber  \\
&&\sum_nU^{\dagger}_{xj,n}U_{xj,n+m}=\delta_{m 0},
\label{Fourier}
\end{eqnarray}
where $\omega_n=2\pi n/\beta, \; n=0, \pm 1, \pm 2,\cdot\cdot \cdot$. 
Then the above correlator is given as 
\begin{equation}
\langle V^{\dagger}_{xj}(\tau_4)V_{xj}(\tau_2)\rangle
=|V_0|^2\sum_{n,m}\langle U^{\dagger}_{xj,n}U_{xj,m}\rangle 
\cdot e^{i(\omega_n\tau_2-\omega_m\tau_4)}.
\label{Vcor}
\end{equation}
From (\ref{Vcor}), it is obvious that if and only if $V_0\neq 0$ and the static mode
of $U_{xj}(\tau)$, i.e., $U_{xj,n=0}$, dominates over all the other oscillating modes,
the quadratic term (\ref{lambda}) gives nontrivial contribution with
the coefficient
\begin{equation}
\langle V^{\dagger}_{xj}(\tau_4)V_{xj}(\tau_2)\rangle
\sim |V_0|^2\langle U^{\dagger}_{xj,0}U_{xj,0}\rangle.
\label{mass}
\end{equation}
Similar term appears from the $\eta_x$-hopping term.
Then the term (\ref{lambda}) behaves as a mass term, and because of that,
the slave-particle constraint becomes less strict at low energies. 
If the above condition for $V_{xj}(\tau)$ is satisfied,  coherent movement of  fluxons $\phi_x$ 
and/or   chargeons $\eta_x$ occur.
As the field $\lambda_x$ can be regarded as a scalar component of the vector potential, the above conclusion implies nothing but the shielding of the static potential, a  phenomena being
observed quite often in many-body systems.

What does the above condition mean for the gauge dynamics ?
In the following sections, we shall show that the above condition is satisfied
if the gauge dynamics of $U_{xj}$ is in the deconfinement phase.
Therefore the above result means that in the deconfinement phase 
the slave-particle constraint is {\em not} faithfully respected by the quasi-excitations.
It is also expected that near the confinement-deconfinement (CD) phase
transition the slave-particle constraint is less effective for quasi-excitations
compared with the original variables.\footnote{Here one should
distinguish the variables for quasi-excitations which are asymptotic fields
from the original variables in the Heisenberg picture, though we often use the same
notation for both of them.
The operator in the Heisenberg picture always satisfies the constraint, but
the asymptotic field does not.
Somewhat detailed discussion on this point is given in the previous 
papers \cite{IMPFS,IMCSS}.}
In the following sections, we shall study the possibility of the PFS, and then
we shall ignore the slave-particle constraint in most of the discussion.

The above conclusion is consistent with the fact that $\lambda_x$ can 
be regarded as a time-component of the transverse gauge field $U_{xj}$.
Actually, the Lagrangian (\ref{LCF}) is invariant under {\em time-dependent} local
gauge transformation with $\lambda_x \rightarrow \lambda_x-\partial_{\tau}\alpha_x$.
Then we can take the temporal gauge, $\lambda_x=\mbox{constant}$, and in this
gauge solely the dynamics of the transverse gauge field $U_{xj}$ determines 
which phase the system is in.
This consideration also supports our treatment of the local constraint in the
subsequent discussions on the phase structure 
of the present model.

In the following subsection, we shall study behavior of the amplitude $V_0$
as a function of $T$.

%%%%%%%%%%%%%%%%%%%%%%%%%%%%%%%%%%%%%%%%%%%%%%%%

\subsection{Amplitude: the hopping expansion}
In this and subsequent subsections, we study behavior of the 
amplitude $V_0$.
We shall use both the hopping expansion and the mean-field type calculation,
which are reliable at intermediate and low $T$, respectively.

Effective potential of $V_0$ is obtained by setting the all link fields as
$V_{xj}=V_0$ in (\ref{effA}), $2 \beta N P(V_0)=-A[V_{xj}=V_0]$,
where $N$ is the total number of sites in the system.
We assume that $V_0$ is a real variable, though this assumption is not
essential for the following calculation.
In this subsection, we shall focus on the single-link potential (SLP) of $V_0$, and
calculate it by the hopping expansion.
From the SLP, we can determine behavior of $V_0$ as a function of $T$.
As we show, the amplitude develops a nonvanishing value at low $T$.

From (\ref{LCF}), at the tree level, we have
\begin{equation}
P_{tree}(V_0)={1 \over 2m}V_0^2.
\label{Ptree}
\end{equation}
In the second order of the hopping expansion, the $\phi_x$ contributes to the 
SLP as 
\begin{eqnarray}
\Delta P^{(2)}_{\phi}  &=& -{1\over (2m)^2}\beta^{-1}V_0^2\int d\tau_1d\tau_2
 \langle \phi^{\dagger}_x\phi_{x+j}(\tau_1)\phi^{\dagger}_{x+j}
\phi_x(\tau_2)\rangle    \nonumber \\
&=&-{1\over (2m)^2}\beta \rho(1-\rho)V_0^2,
\label{P2phi}
\end{eqnarray}
where we have used the  the propagator of $\phi_x$ (\ref{Gphi}).
Similarly, the $\eta_x$ hopping gives contribution as
\begin{eqnarray}
\Delta P^{(2)}_{\eta}&=&-{1\over (2m)^2}\beta^{-1}V_0^2
\int d\tau_1d\tau_2 \langle \eta^{\dagger}_xW_xW^{\dagger}_{x+j}\eta_{x+j}(\tau_1)
\eta^{\dagger}_{x+j}W_{x+j}W^{\dagger}_x\eta_x(\tau_2)\rangle  \nonumber   \\
&=&-{\rho(1-\rho) \over (2m)^2} \beta^{-1}V_0^2
\int d\tau_1d\tau_2 \langle W_xW^{\dagger}
(\tau_1)W_{x+j}W^{\dagger}_x(\tau_2)\rangle.
\label{P2eta1}
\end{eqnarray}
As the fluxon is the hard-core boson, 
the expectation value $\langle WW^{\dagger}WW^{\dagger} \rangle$ in (\ref{P2eta1})
is evaluated by using the following identity which is satisfied for an 
arbitrary c-number $\alpha$
\begin{equation}
e^{\alpha \phi^{\dagger}\phi}=1+(e^{\alpha}-1)\phi^{\dagger}\phi.
\label{hardcore}
\end{equation}
In the leading order of the $\phi$-hopping $\langle WW^{\dagger}WW^{\dagger} \rangle
=1$, and therefore
\begin{equation}
\Delta P^{(2)}_{\eta}=-{\rho(1-\rho) \over (2m)^2}\beta V_0^2.
\label{P2eta}
\end{equation}

Collecting (\ref{Ptree}), (\ref{P2phi}) and (\ref{P2eta}), the quadratic term 
of $V_0$ in the SLP is gives by
\begin{equation}
P^{(2)}(V_0)={1\over 2m}\Big[ 1-{\beta\rho(1-\rho) \over m}\Big]
V_0^2.
\label{P2}
\end{equation}
From (\ref{P2}), it is obvious that, at $T <  T_V \equiv \rho(1-\rho)/m$, 
the amplitude $V_0$ develops a
nonvanishing expectation value.

Higher-order terms of $V_0$ in the SLP is evaluated in a similar way.
In the previous paper \cite{IMPFS}, the quartic term is calculated and $V_0$ is obtained
as a function of $T$.
In the present case, qualitatively same result is obtained, which behaves near $T_V$ as
\begin{equation}
V_0\sim C_V\sqrt{(T_V-T)/T},
\label{V0}
\end{equation}
where $C_V$ is some positive constant (see Ref.\cite{IMPFS} for detailed calculation).

%%%%%%%%%%%%%%%%%%%%%%%%%%%%%%%%%%%%%%%%%%%%%%

\subsection{Amplitude: mean-field calculation at low $T$}

In this subsection, we consider the mean-field type calculation, which is complementary to the hopping expansion in Sect.3.2.
It also shows that $V_0$ develops a nonvanishing expectation value at low $T$.
 
Let us start with the observation that the fluxon  $\phi_x$ field  should 
Bose condense at $T=0$, $\langle \phi_{x} \rangle
  \sim \sqrt{\rho} \exp(i\chi_x)$, due to its 
Bose statistics.($\chi_x$ is the phase of $\phi_x$.)  
Becuase the system is just two-dimensional, the  genuine
long-range order  disappear at $T >0$, but the short-range orders should
survive  well at low $T$. This leads us to a simplification of the Lagrangian
(\ref{LCF}) by  replacing
the nearest-neighbor term $\phi_{x+j}\phi^{\dagger}_x \rightarrow 
\rho \exp(i\chi_{x+j} -i\chi_{x})$.
This also allows us to replace $\phi^{\dagger}_x\phi_x$ in $W_x$ by $\rho$, which implies $W_x \rightarrow 1$. These simplifications give rise to the mean-field Lagrangian $L_{MF}$,
\begin{eqnarray}
L_{\rm MF } &=&-\sum\eta^{\dagger}_x(\partial_{\tau} -\mu_{\eta})\eta_x
\nonumber\\
&&+\sum {V_{0}\over 2m}\Big[(\rho+
\eta^{\dagger}_{x+j}\eta_x)+\mbox{H.c.}\Big] \nonumber\\
&&-2N  {\rho^2\over 2m} 
-\sum{1\over 2m}\Big(\eta^{\dagger}_{x+j}\eta_x\eta^{\dagger}_x\eta_{x+j}\Big)
\nonumber\\
&&-2N {V_0^2\over 2m} 
+ L_{\rm int}(\{\eta^{\dagger}_x\eta_x\}).
\label{LMF}
\end{eqnarray}
 Here we used the unitary gauge $\chi_x =0$.
This represents a system of fermions $\eta_x$ moving with a hopping amplitude
$V_0/(2m)$. To obtain the MF equation that determines the value of $V_0$, 
let us assume that the sum of two four-fermi terms is negligibly small.
This assumption will be explained more in Sect.3.3. Then, in the leading 
order, the system is a collection of  free fermions, and its free 
energy $F_{\rm MF}$ per site can be calculated in a straightforward manner (See Ref.\cite{IMPFS} for details) as
\begin{eqnarray}
\frac{F_{MF}}{N}  &=&      \frac{(V_0-\rho)^2 }{m } 
-\frac{1}{\beta N}\ln \Big[1+  \exp (-\beta  [\frac{V_0}{m} 
\sum_{ i}\cos k_i -\mu_{\eta}]) \Big].
\label{ZMF}
\end{eqnarray}
where $k_i$ is the lattice momentum.
The value of $V_0$ is determined by minimizing $F_{MF}/N$,
\begin{eqnarray}
\frac{\partial}{\partial V_0}\frac{F_{MF}}{N} 
&= & \frac{2(V_0-\rho)}{m} +\frac{1}{m N}\sum_{k } 
    (\sum_{ i}\cos k_i ) f(k) \nonumber\\ 
&=& 0,
\label{MFE}
\end{eqnarray}
where $f(k)$ is the Fermi distribution function,
\begin{eqnarray}
 f(k)& = &
\frac{  \exp\{-\beta 
[\frac{V_0}{m } \sum_{ i}\cos k_i-\mu_{\eta}]\} }
{1+  \exp\{-\beta [\frac{V_0}{m } \sum_{ i}\cos k_i-\mu_{\eta}]\} } .
\label{fdf}
\end{eqnarray}
$\mu_{\eta}$ is determined by the condition for $\rho$,
\begin{eqnarray}
\rho &=& \frac{1}{N}\sum_{k } f(k).
\label{MFrho}
\end{eqnarray}
The set of mean-field equations (\ref{MFE}) and (\ref{MFrho}) 
can be solved numerically. Such an analyses has been done in 
Ref.\cite{IMPFS}. In particular, there is a nonvanishing 
solution for $V_0$ at low $T$. For example, 
$V_0 = \frac{1}{2} +\frac{2}{\pi^2}$ at $T=0$ and $\rho=1/2$.  

%%%%%%%%%%%%%%%%%%%%%%%%%%%%%%%%%%%%%%%%%%%%%%
\subsection{Effective action of  the gauge field $U_{xj}$}

In this subsection, we shall calculate the effective action of the gauge field
$U_{xj}$, $A[U]\equiv A[V_{xj}=V_0U_{xj}]$, and 
in the following section we shall study its phase structure.
We shall employ the hopping expansion, and it gives (approximately) the 
following canonical form of the electric and the magnetic terms of the 
lattice gauge theory:
\begin{equation}
A[U]=A_e+A_m,
\nonumber
\end{equation}
\begin{eqnarray}
A_e&=&-\frac{1}{g^2_{e}}\int d\tau\sum_{x,j} \Big[\partial_{\tau}U^{\dagger}_{xj}\partial_{\tau}U_{xj}+
\cdot\cdot\cdot\Big]   \nonumber  \\
A_m&=&\frac{1}{g^2_{m}}\int d\tau \sum_x\Big[U_{x,2}U_{x+2,1}U^{\dagger}_{x+1,2}U^{\dagger}_{x,1}
+\mbox{H.c.}+\cdot\cdot\cdot\Big],
\label{AeAm}
\end{eqnarray}
where ${g^2_{e}}$ and ${g^2_{m}}$ are the effective electric and magnetic 
gauge couplings, respectively.
The reader who is not interested in the detailed derivations given  below
may skip them; just note the main result given in Eqs.(\ref{Ae}) and below, 
and go to Sect.4 to find the  discussion on the physical results like 
the phase structure, etc. 

The $\phi-$ and $\eta-$hopping terms and the $\phi^4$ and the $\eta^4$ 
interactions
in (\ref{LCF}) give each contribution to the effective action,
which we write as 
\begin{eqnarray}
A_e&=&A_{e,\phi}+A_{e,\eta}+A_{e,\phi^4}+A_{e,\eta^4},  \nonumber   \\
A_m&=&A_{m,\phi}+A_{m,\eta}+A_{m,\phi^4}+A_{m,\eta^4}.
\label{Aeach}
\end{eqnarray}  

Let us start with $A_{e,\phi}$. As in the calculation of the SLP in Sect.3.2, 
in the second-order of the hopping expansion of $\phi_x$, we  have
\begin{eqnarray}
A_{e,\phi}&=&|V_0|^2{1\over (2m)^2}\sum\int d\tau_1 d\tau_2
U^{\dagger}_{xj}(\tau_1)
U_{xj}(\tau_2)\langle \phi^{\dagger}_x\phi_{x+j}(\tau_1)\phi^{\dagger}_{x+j}
\phi_x(\tau_2)\rangle  \nonumber  \\
&=&|V_0|^2{1\over (2m)^2}\rho(1-\rho)\beta^2\sum U^{\dagger}_{xj,0}U_{xj,0},
\label{Aephi}
\end{eqnarray}
where $U_{xj,0}$ is the static component of the Fourier decomposition 
of $U_{xj}$(\ref{Fourier}).

Similarly, the $\eta_x$ hopping gives rise to $A_{e,\eta}$ as 
\begin{eqnarray}
A_{e,\eta}&=&|V_0|^2{1\over (2m)^2}\sum\int d\tau_1 d\tau_2U^{\dagger}_{xj}(\tau_1)
U_{xj}(\tau_2)\langle \eta^{\dagger}_xW_xW^{\dagger}_{x+j}\eta_{x+j}(\tau_1)
\eta^{\dagger}_{x+j}W_{x+j}W^{\dagger}_x\eta_x(\tau_2)\rangle  \nonumber  \\
&=&|V_0|^2{\rho(1-\rho)\over (2m)^2} \beta^2\sum  U^{\dagger}_{xj,0}U_{xj,0}.
\label{Aeeta}
\end{eqnarray}

Both $A_{e,\phi}$ and $A_{e,\eta}$ above 
have the form $\sum U^{\dagger}_{xj,0}U_{xj,0}$. From the factor $\beta^2$
in  (\ref{Aephi}) and (\ref{Aeeta}), it is obvious that, at high $T$, the
coefficients of these terms are small and almost no significant enhancement or
depression  appear for $U_{xj}$'s; All the modes $U_{xj,n}$ fluctuate randomly.
On the other hand,  at low $T$, the coefficients develop and 
the static mode $U_{xj, n=0}$ dominates over all the other oscillating modes.
By using the unitarity condition (\ref{Fourier}), these electric terms
are rewritten as the follwing canonical form effectively,
\begin{eqnarray}
\beta^2 U^{\dagger}_{xi,0}U_{xi,0}&=&\beta^2\Big(1-\sum_{n\neq 0}
U^{\dagger}_{xi,n}U_{xi,n}\Big)  \nonumber  \\
&\sim& \beta^2-{2 \beta^3\over (2\pi)^2}\int^{\beta}_0 
d\tau\partial_{\tau}U^{\dagger}_{xi}
\partial_{\tau}U_{xi}.
\label{rewrite1}
\end{eqnarray}
Therefore, the effective gauge coupling $g^2_e$ in this system 
has strong $T$ dependence. Because of this fact, the present gauge system 
exhibits a rather nontrivial phase structure
as $T$ changes.

Let us turn to the contributions from the interaction terms in
$L_{\eta \phi V}$ of  (\ref{LCF}) with $L_{\rm int}$ of (\ref{inter}).
It is not so easy to evaluate their effects for general coupling constants $g_1$ and $g_2$.
Therefore, we assume that the interaction between fermions almost
cancels the $\phi^4$ and the $\eta^4$ terms in $L_{\eta\phi}$  
which appear as a result of the introduction of the ``gauge field" $V_{xj}$.
That is, we adjust the parameters $g_1$ and $g_2$ such that the total results,
\begin{eqnarray}
&&{1\over 2m}\Big(\phi^{\dagger}_{x+j}\phi_{x+j}\phi^{\dagger}_x\phi_x\Big)
+{1\over 2m}\Big(\eta^{\dagger}_{x+j}\eta_{x+j}\eta^{\dagger}_x\eta_x\Big)
+L_{\rm int}(\{\eta^{\dagger}_x\eta_x,\phi^{\dagger}_x\phi_x\})  \nonumber  \\
&& \; \; =
\lambda_1\Big(\eta^{\dagger}_{x+j}\eta_{x+j}\eta^{\dagger}_x\eta_x\Big)
+\lambda_2\Big(\phi^{\dagger}_{x+j}\phi_{x+j}\phi^{\dagger}_x\phi_x\Big), \label{4int} 
\end{eqnarray}
have  small constants $\lambda_1 (= (2m)^{-1} + g_1)$ and $\lambda_2
(=  (2m)^{-1} + g_2)$. 
We shall use the perturbative 
calculation in powers of these $\lambda_1$ 
and $\lambda_2$.
This calculation still give  some important results, as we shall see.
This assumption implies  $g_1 < 0$ and $g_2 < 0$, that is, there
 exist repulsions between 
fluxons and also between chargeons.
For electrons in the half-filled Landau level, only chargeons have EM charge, 
$g_1 < 0$ and $g_2 = 0$, so $\lambda_2= (2m)^{-1}$.

Explicitly, for  $A_{e,\phi^4}$, the $\phi^4$ term is contracted with
the fluxon hopping, giving rise to 
\begin{eqnarray}
A_{e,\phi^4}&=&{\lambda_2\over (2m)^2}|V_0|^2
\int \prod_{i=1}^3d\tau_i U^{\dagger}_{xj}(\tau_1)
U_{xj}(\tau_2)\langle \phi^{\dagger}_x(\tau_1)
\phi_{x+j}(\tau_1)\phi^{\dagger}_{x+j}(\tau_2)
\phi_x(\tau_2)  \nonumber  \\
&& \times \;(\phi^{\dagger}_{x+j}(\tau_3)\phi_{x+j}(\tau_3)-\rho)
(\phi^{\dagger}_x(\tau_3)\phi_x(\tau_3)-\rho) \rangle \nonumber   \\
&=&{\lambda_2\over (2m)^2}|V_0|^2\beta^3\rho^2(1-\rho)^2\sum U^{\dagger}_{xj,0}
U_{xj,0}.
\label{Aephi4}
\end{eqnarray}
Here  we have replaced the $\phi^4$ term (and $\eta^4$ term) by its  
``normal-ordered" form, 
$(\phi^{\dagger}_{x+j}\phi_{x+j}-\rho)
(\phi^{\dagger}_x\phi_x-\rho) $ for convenience. This brings an irrelevant
 shift of  chemical potential and addition of an irrelevant constant to the Lagrangian.

Similarly, for $A_{e,\eta^4}$ we have
\begin{eqnarray}
A_{e,\eta^4}&=&{\lambda_1\over (2m)^2}
|V_0|^2\int \prod_{i=1}^3d\tau_i U^{\dagger}_{xj}(\tau_1)
U_{xj}(\tau_2)\nonumber\\
&&\langle \eta^{\dagger}_x(\tau_1)W_xW^{\dagger}_{x+j}\eta_{x+j}(\tau_1)
\eta^{\dagger}_{x+j}(\tau_2)W_{x+j}W^{\dagger}_x
\eta_x(\tau_2)  \nonumber  \\
&& \times \;(\eta^{\dagger}_{x+j}(\tau_3)\eta_{x+j}(\tau_3)-\rho)
(\eta^{\dagger}_x(\tau_3)\eta_x(\tau_3)-\rho) \rangle \nonumber   \\
&=&-{\lambda_1 \over (2m)^2}|V_0|^2\beta^3
\rho^2(1-\rho)^2\sum U^{\dagger}_{xj,0}U_{xj,0}.
\label{Aeeta4}
\end{eqnarray}
Collecting these terms, we obtain the result for $A_{e}$, 
\begin{eqnarray}
&&A_{e,\phi}+A_{e,\eta}+A_{e,\phi^4}+A_{e,\eta^4}  \nonumber  \\
&& \; \; \; \sim -{1\over (2m)^2}|V_0|^2\rho(1-\rho){\beta^3 \over (2\pi)^2}
\Big[2+(\lambda_2-\lambda_1)\rho(1-\rho)\beta\Big]  \nonumber   \\
&& \;\; \; \; \times \; \sum \int d\tau \partial_{\tau}U^{\dagger}_{xi}
\partial_{\tau}U_{xi}.
\label{Ae}
\end{eqnarray}
Thus the effective electric gauge-coupling constant $g^2_{e}$ is given by 
\begin{equation}
g^2_{e}=\Bigg(|V_0|^2{\rho(1-\rho)\over (2m)^2}{\beta^3 \over (2\pi)^2}
\Big[2+(\lambda_2-\lambda_1)\rho(1-\rho)\beta\Big]\Bigg)^{-1}.
\label{geff}
\end{equation}

The magnetic terms are calculated in a similar way.
They determine spatial configuration of the gauge field $U_{xj}$.
From the fluxon hopping, we have
\begin{eqnarray}
A_{m,\phi}&=&\Big({V_0 \over 2m}\Big)^4\sum_x\prod^4_{i=1}\int d\tau_i
\Big(U_{x2}(\tau_4)U_{x+2,1}(\tau_3)U^{\dagger}_{x+1,2}(\tau_2)U^{\dagger}_{x1}
(\tau_1)+\mbox{H.c.}\Big)  \nonumber   \\
&&\; \times \; \prod^4_{i=1}G_{\phi}(\tau_i-\tau_{i+1})  \nonumber  \\
&\simeq&C_{\phi}\Big({V_0\over 2m}\Big)^4\beta^4\sum_x
\Big(U_{x2,0}(\tau_4)U_{x+2,1,0}(\tau_3)U^{\dagger}_{x+1,2,0}(\tau_2)U^{\dagger}_{x1,0}
(\tau_1)+\mbox{H.c.}\Big),
\label{Amphi}
\end{eqnarray}
where $C_{\phi}$ is given by
\begin{equation}
C_{\phi}={1\over 4!}\Big\{4\rho(1-\rho)^3+12\rho^2(1-\rho)^2+4\rho^3(1-\rho)\Big\}.
\label{Cphi}
\end{equation}
In the last line of (\ref{Amphi}), we have retained only the terms of products of four
zero modes $U_{xj,0}$.
Similarly, from the $\eta_x$-hopping, we get
\begin{eqnarray}
A_{m,\eta}&=&-\Big({V_0\over 2m}\Big)^4\sum_x\prod^4_{i=1} \int d\tau_i
\Big( U_{x2}(\tau_4)U_{x+2,1}(\tau_3)
U^{\dagger}_{x+1,2}(\tau_2)U^{\dagger}_{x1}(\tau_1)
 \Big)   \nonumber  \\
&&\times \prod^{4}_{i=1}G_{\eta}(\tau_i-\tau_{i+1})  \nonumber  \\
&& \times \langle W^{\dagger}_xW_{x+1}W^{\dagger}_{x+1}
W_{x+1+2}W^{\dagger}_{x+1+2}W_{x+2}W^{\dagger}_{x+2}
W_x\rangle +\mbox{H.c.}   \nonumber  \\
&=& -\Big({V_0\over 2m}\Big)^4
\sum_x\prod^4_{i=1} \int d\tau_i
 \Big( U_{x2}(\tau_4)U_{x+2,1}(\tau_3)
U^{\dagger}_{x+1,2}(\tau_2)U^{\dagger}_{x1}(\tau_1)
 \Big)   \nonumber  \\
&& \times \prod^{4}_{i=1}G_{\eta}(\tau_i-\tau_{i+1}) \;
e^{-2\pi qi\rho}  \langle e^{2\pi qi{\phi}_{x}^{\dagger}\phi_{x}}
\rangle +\mbox{H.c.} \nonumber  \\
&\simeq& -C_{\eta}\Big({V_0\over 2m}\Big)^4 \beta^4   
\sum_x \Big( U_{x2,0}U_{x+2,1,0}
U^{\dagger}_{x+1,2,0}U^{\dagger}_{x1,0} \Big) \nonumber  \\
&&\times e^{-2\pi iq\rho}[1-\rho +e^{2\pi iq } \rho] +\mbox{H.c.},
\nonumber\\
C_{\eta}&= &{1\over 4!}\Big\{-4\rho(1-\rho)^3+
12\rho^2(1-\rho)^2-4\rho^3(1-\rho)\Big\}.
\label{Ameta}
\end{eqnarray}
where we have used  $\nabla_i \nabla_i G(x,x') = \delta_{x x'}$. 

For the case of the half-filled Landau level, one sets $\rho=1/2$ and $q=2$,
which leads to $C_{\phi}>C_{\eta}$.
Therefore, the lowest-energy state is realized by the fluxless and uniform 
configuration of $U_{xj}$'s; $\langle U_{xj} \rangle$ = constant.
There are also contributions from the interaction terms $A_{m,\phi^4}$ and 
$A_{m,\eta^4}$.
Anyway, it is true that the magnetic term $A_m$ also prefers the static modes 
of the gauge field at low $T$.

%%%%%%%%%%%%%%%%%%%%%%%%%%%%%%%%%%%%%%%%%%%%%
\setcounter{equation}{0}
\section{Phase structure of the effective gauge theory and quasi-particles}

Before discussing the PFS in our effective gauge model derived in Sect.3.4, 
let us recall some general arguments on the CD transition of the
 canonical lattice gauge theory, whose effective gauge coupling constant
 $g^2_e = g^2_m = g^2_{can}$ is $T$-independent.
From the work of Polyakov and Susskind \cite{PS} and the explicit 
Monte Carlo simulations, it is well-known that the 
CD phase transition takes place
at finite $T$  for such a canonical lattice gauge theory. In Ref.\cite{PS},
this CD phase transition is observed by mapping the strongly-coupled gauge
system to an effective classical spin model \cite{PS}; the CD transition is 
identified with the order-disorder transition of the spin dynamics.
For compact U(1) gauge theory in $(2+1)$ dimensions, 
the mapped spin model is the XY model in two dimensions which 
exhibits the Kosterlitz-Thouless (KT) phase transition.\footnote{
For a more complicated effective gauge theory derived for  the t-J model, 
the mapping is performed in Ref.\cite{IMCSS}.} 
According to Ref.\cite{PS}, for the canonical gauge system, 
the transition temperature $T^{\ast}_{\rm CD}$ is estimated as 
\begin{equation}
T^{\ast}_{\rm CD} \simeq g_{can}^2.
\label{TCD}
\end{equation}
It is concluded that the  deconfienement phase appears at  
$T>T^{\ast}_{\rm CD}$,
while the confinement phase appears at  $T < T^{\ast}_{\rm CD}$, .
This result is easily seen in the Largangian formalism.
In terms of the Fourier components of the gauge field (\ref{Fourier}),
the action of the canonical gauge system is written as 
\begin{eqnarray}
A_{can}&=&-{1\over g_e^2}\int d\tau \sum_{x,j} 
\partial_{\tau}U^{\dagger}_{xj}(\tau)
\partial_{\tau}U_{xj}(\tau)  \nonumber  \\
&=&-{\beta \over g_e^2}\sum_{x,j}\sum_n \omega_n^2 U^{\dagger}_{xj,n}
U_{xj,n},
\label{Acan}
\end{eqnarray}
where $\omega_n=2\pi n/\beta$.\footnote{Of course the action contains also the magnetic term.
In the usual consideration \cite{PS} its effect is neglected preferring a simple treatment.  
It is shown   that the magnetic term enhances the deconfinement phase, hence the existence 
of the CD  transition  itself remains true.}
Then it is obvious that at very high $T$, fluctuations of 
all the oscillating modes $U_{xj,n\neq 0}$ are suppressed and only the static 
mode $U_{xj,n= 0}$
develops its amplitude.
On the other hand,  at very low $T$, the oscillating modes as well as 
the static mode fluctuate randomly.
There should be a CD phase transition at some intermediate $T$, that is
at  $T=T^{\ast}_{\rm CD}$.

Now let us consider 
the effective gauge model derived in Sect.3.4 for 
the CS gauge theory coupled with nonrelativistic fermions.
The effective gauge couplings $g^2_{e}$ and $g^2_m$ have 
strong-$T$ dependence as given by
(\ref{geff}), in contrast with the usual canonical lattice gauge theory.
This is bacause  the present gauge field $U_{xj}$ is not a genuine 
gauge field, but is a composite, or ``bound state"
of the ``elementary fields" $\phi_x$ and $\eta_x$.
It is obvious that at low $T$, $g^2_{e} \rightarrow$ 0 very rapidly,
and at high $T$, $g^2_{e} \rightarrow \infty$.
Therefore we expect again a CD transition.
Actually, from the explicit form of (\ref{geff}) and (\ref{TCD}), 
we conclude that the CD phase
transition occurs at 
\begin{equation}
T_{\rm PFS}\simeq \Big({\rho(1-\rho) \over 2m^2} \Big)^{1/2}
{V_0\over 2\pi}+{1\over 4}\rho(1-\rho)(\lambda_2-\lambda_1).
\label{TCD2}
\end{equation}
Due to the strong $T$-dependence of $g^2_e$, the confinement phase
takes place at $T$ {\it above} $T_{\rm PFS}$, and
the {\em deconfinement}
phase, i.e., the PFS, takes place at $T$ {\it lower} than $T_{\rm PFS}$.
To be able to neglect the higher-order terms, the parameter $\lambda_1 = (2m)^{-1} + g_1$ in (\ref{4int}) must be small, which implies that
the coefficient $g_1$  of the  {\em repulsion} between EM charges should be
 about  $g_1 \sim -(2m)^{-1}$. When the repulsion become stronger, $\lambda_1$ become negative and $T_{\rm PFS}$ rises, that is, the PFS is more enhanced.
As explained before, this result supports the intuitive physical expectation
for the stability of CF's in the half-filled Landau level.
This point can be rephrased as follows; If there were no repulsions between
EM charges at all, the assumption of smallness of $\lambda$'s would lose its 
support. Then there might be  no convincing calculations showing that the PFS still takes place.
Actually it is quite possible that the PFS disappears at certain point as
the strength of the repulsion is decreased.\footnote{The extreme case of this 
possibility is the system of {\it free} electrons without repulsions. One may
rewrite the system by some constituent operators. But we know that the
system cannot exhibit any   separation phenomena at all.} 

In (\ref{TCD2}), the amplitude $V_0$ itself depends on $T$.
Actually, it is a decreasing function of $T$ 
and so  Eq.(\ref{TCD2}) has a unique solution for $T_{\rm PFS}$.
We note that the expression (\ref{TCD2}) is different from the
corresponding expression calculated in Ref.\cite{IMPFS}, reflecting
the different treatments of the flux degrees of freedom.
For the t-J model based on the slave-boson formalism, we performed 
systematic numerical calculations for the mean-field amplitudes and 
determined the transition $T$ of CSS, $T_{\rm CSS}$, at each hole 
concentration \cite{IMS}.
The result shows that the effect of gauge-field fluctuations is so large
that $T_{\rm CSS}$ is reduced to about 10 $\%$
of the mean-field critical temperature $T_{\rm V}$ which is defined
as the $T$ at which the mean field
$V_0$ vanishes as in (\ref{V0}); thus we have  $T_{\rm CSS}\sim {1\over 10}T_{V}$.
We can expect similar behavior for the PFS.
Practical calculations of $T_{\rm PFS}$ and the effective mass of the CF
is under study and results will be reported in a forthcoming paper.
There we shall also compare $T_{\rm PFS}$ numerically with that 
of Ref.\cite{IMPFS}

Let us discuss the nature of each phase in some detail.
In the confinement phase, as explained in Sect.2, 
the gauge field fluctuates so strongly that
only charge-neutral compounds with respect to
(\ref{gaugetrf}) and (\ref{gaugetrfV}) appear as quasi-excitations.
They are bound states of the chargeon and the fluxon like $\eta_x\phi_x$, 
$\phi_{x+j}U_{xj}\eta_x$, etc.\footnote{Of course, 
only the low-lying energy states of
 linear combinations of these bound states appear  as quasi-particles.}
In the system of the half-filled Landau level, they are nothing but the 
original electrons, which are bound state of the CF and flux quanta.
In this phase, the CF's are {\em not}  quasi-excitations.

On the other hand, in the deconfinement phase of $U_{xj}$'s, 
the fluctuations of gauge field are small,
so the quasi-particles carry the same quantum numbers as 
the ``elementary" fields that appear in the Hamiltonian and couple 
with the gauge field.
Therefore in the present system, the quasi-particles are the chargeons, 
the fluxons and the gauge bosons. We stress that 
the gauge field itself is an independent 
degree of freedom
and appears as quasi-excitations.
Actually, the transverse components of the gauge field is {\em not} 
shielded in contrast with the longitudinal part which is shielded as we have
seen in Sect.3.1.
The transverse gauge field  produces nontrivial effects on 
the  chargeons and fluxons at low energies.

To understand this mechanism, it is convenient to introduce 
the following variable 
$\tilde{V}_{xj}$ instead of $V_{xj}$,
\begin{equation}
\tilde{V}_{xj}=V_{xj}W_{x+j}W^{\dagger}_x.
\label{tilV}
\end{equation}
Then the hopping terms of chargeons and fluxons in (\ref{LCF}) 
are written as 
\begin{equation}
 \phi^{\dagger}_{x+j}\tilde{V}_{xj}W^{\dagger}_{x+j}W_x\phi_x
+\eta^{\dagger}_{x+j}\tilde{V}_{xj}\eta_x+\mbox{H.c.}
\label{hopping}
\end{equation}
The field $\tilde{V}_{xj}$  as well as $V_{xj}$  
can be regarded as a {\it dynamical} gauge field,
which is generated as a result of the PFS.
If there were {\em no} $\phi$-hopping term, the generated gauge field 
$\tilde{V}_{xj}$ 
had no correlations with the CS gauge field which represents fluxes 
attaching fermions;
The chargeons $\eta_x$ would be free from any constraints and just move 
interacting with that ``gauge field"  
whose kinetic term does not exist at the tree level.
In the random-phase approximation (RPA) or the 
renormailzation-group study (RGS),
a kinetic term of this gauge field appears from the loop effects of matter 
fields, i.e., chargeon in the present case.
Recently, related  models of gauge theory coupled with nonrelativistic fermions are studied \cite{RG}.
It has been shown that fermions exhibit non-Fermi-liquid behavior like 
the Luttinger liquid in (1+1)
dimensions or the marginal Fermi liquid in high-T$_C$  cuprates.

However, in the present case, there exists the first term in (\ref{hopping});
  the gauge field $\tilde{V}_{xj}$ {\it does} interact also 
with the fluxons, and this flux-hopping term gives rise to 
the correlation between 
the dynamical gauge field $\tilde{V}_{xj}$ and the CS gauge field, i.e., 
\begin{equation}
\tilde{V}_{xj} \sim W^{\dagger}_xW_{x+j}.
\nonumber
\end{equation}
In the recent studies of the gauge theory of fermions in the half-filled Landau level \cite{RG,RPA,IMO}, one first starts with the CS gauge theory.
However, in the RPA or the RGS, the CS constraint is totally ignored.
This is essentially due to the technical difficulty in
 handling the CS constraint.  It should be noted that even in the PFS state,
 the  CS constraint is to be {\em partially} respected by the real 
 quasi-excitations; the chargeons and fluxons necessarily interact
 via the gauge field. Such effects are important for quantitative analyses.
 In the present system, just the fluxons are in charge of the correlations
  between the dynamical gauge field and the CS fluxes.
Then, it is very interesting to study the present gauge theory of 
chargeons and fluxons by the RPA and/or the RGS (in the continuum) 
at {\it low or zero} $T$.
The local constraint (\ref{cons}) can be partially taken into account 
   through the {\em massive}
Lagrange multiplier $\lambda_x$.
Such analyses are complementary to the studies in this paper by 
the hopping expansion
on the lattice. The hopping expansion assume the smallness of 
$\langle U_{xj} \rangle$, 
the order parameter of PFS, hence  reliable at $T$ near $T_{\rm PFS}$.
As an example of the importance of such a residual correlation effect,
we have a mass of CF. As explained before, the
 chargeon is the CF in the case of half-filled Landau level.
Its mass read off from the Lagrangian is given by 
\begin{equation}
m_{\rm CF}= \frac{m}{V_0},
\label{massCF}
\end{equation}
and  we estimated as $V_0 = $constant of $O(1)$ at low $T$ in Sect.3.3.
There are various kinds of  experiments \cite{exper,exper2}
measuring $m_{\rm CF}$ through different physical quantities.
The estimated values of $m_{\rm CF}$ are scattered in a spectrum, 
$m_{\rm CF}/m = O(1) \sim O(10)$.
These results should be coherently explained by the 
residual correlation effects.

%%%%%%%%%%%%%%%%%%%%%%%%%%%%%%%%%%%%%%%%%%%%%%%
\setcounter{equation}{0}
\section{Conclusion}

In this paper, we have studied the CS gauge theory of nonrelativistic fermions.
Especially, we are interested in the phenomenon of the PFS.
This is very important for the CF approach to the electrons in 
the half-filled Landau level
and also for the FQHE {\it near the half filling} \cite{IMO}.
The main problem of the CS gauge theory is how to treat the CS constraint.
To this end, we have introduced the chargeon and the fluxon operators, 
and expressed
the original fermion operator by a bilinear form of them.

To discuss the PFS,
we have rewritten the model by using the gauge field, which glues the chargeon
and the fluxon.
It is shown that the possibility of the PFS or (ir)relevance 
of the CS constraint is reduced to   the possibility of the  
CD phase transition of that gauge field.
We showed by using the hopping expansion that the PFS takes place at low
$T < T_{\rm PFS}$. 
The repulsion between charges plays a very important role for PFS.
This result supports the intuitive physical expectation fot
 the stability of CF's.

In contrast to the previous approach \cite{IMPFS}, the present approach
has a new operator, the fluxon $\phi_x$, which opens a possibility to 
describe the higher-order effects of CS constraint, such as the mass of CF
near the half filling. More generally, it is certainly an improvement that
a field-theoretical description is now possible for the dynamics of CS flux
degrees of freedom as well as of CS charge degrees of freedom.
Also, as pointed out,  the chargeon-fluxon approach  reveals  the strong
resemblance to the slave-boson or fermion approach to the t-J model.
This open a possibility of a more coherent and universal understanding of 
these separation phenomena. We will return to these topics in  future.

%%%%%%%%%%%%%%%%%%%%%%%%%%%%%%%%%%%%%%%%%%%%%%

\eject
						
\begin{tabular}{|r|c|c|c|c|}  \hline                 & Flux quanta    & CS charge   &  $V$ charge & EM charge   \\\hline
electron $ C^{\dagger}_x  $   &    $0 $ &  $0$       &    0 & $-e$   
\\\hline
fermion  $ \psi^{\dagger}_x$   &    $q $  &  $1$  &  0 &  $-e$     
\\\hline
chargeon $ \eta^{\dagger}_x$   &  $ 0$    &  $1$   &  1 &   $-e$                  \\\hline
fluxon $ \phi^{\dagger}_x$   &   $q$   &   $0$  &  $-1$ &  $0$              
\\ \hline
\end{tabular}
\\

Table 1. Quantum numbers carried by  various elementary fields, $C^{\dagger}_x$, $\psi^{\dagger}_x ( = \eta^{\dagger}_x \phi^{\dagger}_x)$, 
 $\eta^{\dagger}_x$,  $\phi^{\dagger}_x.$ In the confinement phase 
of the gauge dynamics of $V_{xj}$ at $T > T_{\rm PFS}$,
only the neutral objects of V charge, like $\psi_x$ ($C_x$), appear as physical excitations.
In the deconfinement phase at $T < T_{\rm PFS}$, the PFS is realized and 
V-charged objects, like $\eta_x, \phi_x $,  can apper. When fluxons $\phi_x$ Bose-condense, $\phi_x \simeq \sqrt{\rho}$ (at $T=0$),  
two candidates for CF's,  $\psi^{\dagger}_x$ and $\eta^{\dagger}_x$, 
become indistinguishable in the leading order.

\eject
Figure Captions.

{\bf Fig.1:}
Illustrations of  the key concepts and objects appeared 
in each step to reach $L_{\eta\phi V}$ of (\ref{LCF}). 

{\bf Fig.1a:}
Illustration of a system of electrons under an external magnetic field 
$B^{\rm ex}$ in the $z$-direction. Each black bullet represents an electron,
and straight lines with arrows represent  $B^{\rm ex}$.
  
{\bf Fig.1b:} 
Illustration of Eq.(\ref{CF}).
Each electron $C^{\dagger}_x $ is represented as a product of a fermion
 $\psi^{\dagger}_x$ and the operator $\tilde{W}^{\dagger}_x \equiv \exp[-iq\sum\theta(x-y)\hat{\rho}_y]$. The latter represents $q$ units of 
CS fluxes in the {\it negative} $z$-direction, which are represented by 
wavy lines with arrows. The sources of these fluxes are the 
fermions themselves as shown in Eq.(\ref{BCS}). 

{\bf Fig.1c:} Illustration of Eq.(\ref{bilinear}).
Each fermion $\psi^{\dagger}_x$ is a compoiste of
a fluxon $\phi^{\dagger}_x$ and a chargeon $\eta^{\dagger}_x$.
A broad  arrow represents a  fluxon which carries $q$ units of CS fluxes
in the $z$-direction, the  sources of which are  nothing but the fluxons themselves
as shown in Eq.(\ref{BCS2}).
An open circle represents a chargeon which carries CS  and EM charges.
When the PFS phenomenon takes place, $\psi^{\dagger}_x$ dissociates into
$\phi^{\dagger}_x$ and  $\eta^{\dagger}_x$. 

{\bf Fig.1d:}
Illustration of the system of Fig.1a in the chargeon-fluxon representation.
As seen from Eq.(\ref{LCF}), the chargeons move in the  statistical potential $\Delta A_{xj} \equiv eA_{xj}^{ex} - A_{xj}^{CS}$ described by $W_{xj}$
and, at the same time,  in the gauge field $V_{xj}$.( $V_{xj}$ has not
been drawn explicitly.)  We depicted
 the magnetic field $\Delta B$ corresponding to this $\Delta A_{xj}$ by 
 black broad arrows, and put them on each fluxon since its local part 
$B^{\rm CS}$ sits at each fluxon as shown by Eq.(\ref{BCS2}). The fluxons 
have minimal interactions with $V_{xj}$. In the PFS state, cancellation
between $eA^{\rm ex}$ and $A^{\rm CS}$ works well (e.g.,  perfectly in the
leading order of Bose condensation, $\phi_x^{\dagger} \phi_x = \rho$),
 and the fluctuation effects by
$\Delta A_{xj}$ can be treated legitimately as a perturbation.\\


\newpage
\begin{thebibliography}{1}
\bibitem{Girvin}S.M.Girvin and A.H.MacDonald, 
{\it Phys.Rev.Lett.}{\bf 58}(1987)1252;  \\
S.C.Zhang, T.H.Hansen and S.Kivelson, 
{\it Phys.Rev.Lett.}{\bf 62}(1989)82; \\
N.Read, {\it Phys.Rev.Lett.}{\bf 62}(1989)86.
%
\bibitem{Jain}J.Jain,
 {\it Phys.Rev.Lett.}{\bf 63}(1989)199;
{\it Phys.Rev.}{\bf B40}(1989)8079; {\bf B41}(1990)7653.
%
\bibitem{exper}See for example, \\
H.C.Manoharam, M.Shayegan and S.J.Klepper, 
{\it Phys.Rev.Lett.}{\bf 73}(1994)3270; \\
R.R.Du, H.L.Stormer, D.C.Tsui, A.S.Yeh, L.N.Pfeiffer and K.W.West, \\
{\it Phys.Rev.Lett.}{\bf 73}(1994)3274.
%
\bibitem{RR}E.Rezayi and N.Read,
{\it Phys.Rev.Lett.}{\bf 72}(1994)900.
%
\bibitem{IMO}
I.Ichinose, T.Matsui, M.Onoda,
{\it Phys.Rev.}{\bf B52}(1995)10547.
%
\bibitem{Anderson}P.W.Anderson,
{\it Phys.Rev.Lett.}{\bf 64}(1990)1839.
%
\bibitem{IMPFS}I.Ichinose and T.Matsui,
{\it Nucl.Phys.}{\bf B468[FS]}(1996)487.
% 
\bibitem{IMCSS}I.Ichinose and T.Matsui, 
{\it Nucl.Phys.}{\bf B394}(1993)281; \\
{\it Physica} {\bf C235-240}(1994)2297; 
{\it Phys.Rev.}{\bf B51}(1995)11860.
%
\bibitem{RG}C.Nayak and F.Wilczek,
{\it Nucl.Phys.}{\bf B417[FS]}(1994)359; {\bf B430[FS]}(1994)534;  \\
S.Chakravarty, R.E.Norton and O.F.Syljuasen,
{\it Phys.Rev.Lett.}{\bf 74}(1995)1423;  \\
I.Ichinose and T.Matsui, {\it Nucl.Phys.}{\bf B441[FS]}(1995)483;\\
M.Onoda, I.Ichinose and T.Matsui, 
{\it Nucl.Phys.}{\bf B446[FS]}(1995)353.
%
\bibitem{PS}
A.M.Polyakov,
{\it Phys.Lett.}{\bf B72}(1978)477; \\
L.Susskind,
{\it Phys.Rev.}{\bf D20}(1979)2610.
%
\bibitem{RPA}B.I.Halperin, P.A.Lee and N.Read,
 {\it Phys.Rev.}{\bf B47}(1993)7312;  \\
J.Polchinski,
{\it Nucl.Phys.}{\bf B422}(1994)617;  \\
B.L.Altshuler, L.B.Ioffe and A.J.Millis,
{\it Phys.Rev.}{\bf B50}(1994)14048.
%
\bibitem{exper2}W.Kang, S.He, H.L.Stormer, L.N.Pfeiffer, K.W.Baldwin and K.W.West,  \\
{\it Phys.Rev.Lett.}{\bf 75}(1995)4106.
%
\bibitem{IMS}I.Ichinose, T.Matsui and K.Sakakibara,
preprint UT-Komaba 96-10,
{\it `` Fluctuation effects of gauge fields in the slave-boson t-J model."}




 
 
\end{thebibliography}
\end{document}